\documentclass{PoS}

\title{The beam energy dependence of collective flow in heavy ion collisions}

\ShortTitle{Energy dependence of collective flow}

\author{\speaker{Hannah Petersen}\\
        Institut f\"ur Theoretische Physik, Johann Wolfgang Goethe-Universit\"at, Max-von-Laue-Str. 1, D-60438 Frankfurt am Main, Germany\\
        Frankfurt Institute for Advanced Studies, Ruth-Moufang-Str. 1, D-60438 Frankfurt am Main, Germany\\
        GSI Helmholtzzentrum f\"ur Schwerionenforschung GmbH, Planckstr. 1, 64291 Darmstadt, Germany\\
        E-mail: \email{petersen@th.physik.uni-frankfurt.de}}

\author{Jussi Auvinen\\
         Department of Physics, Duke University, POBOX 90305, Durham, NC 27707, USA\\
         E-mail: \email{auvinen@fias.uni-frankfurt.de}}

\author{Jan Steinheimer\\
		Frankfurt Institute for Advanced Studies, Ruth-Moufang-Str. 1, D-60438 Frankfurt am Main, Germany\\
		E-mail: \email{steinheimer@fias.uni-frankfurt.de}}

\author{Marcus Bleicher\\
		Institut f\"ur Theoretische Physik, Johann Wolfgang Goethe-Universit\"at, Max-von-Laue-Str. 1, D-60438 Frankfurt am Main, Germany\\
        Frankfurt Institute for Advanced Studies, Ruth-Moufang-Str. 1, D-60438 Frankfurt am Main, Germany\\
        E-mail: \email{bleicher@th.physik.uni-frankfurt.de}}

\abstract{The major goals of heavy ion research are to explore the phase diagram of quantum chromodynamics (QCD) and to investigate the properties of the quark gluon plasma (QGP), a new state of matter created at high temperatures and/or densities. Collective anisotropic flow is one of the most promising observables to gain insights about the properties of the system created in relativistic heavy ion reactions. The current status of the beam energy dependence of the first three Fourier coefficients of the azimuthal distribution of the produced particles $v_1$ to $v_3$ within hybrid transport plus hydrodynamics approaches are summarized. }

\FullConference{9th International Workshop on Critical Point and Onset of Deconfinement\\
		 17-21 November, 2014\\
		 ZiF (Center of Interdisciplinary Research), University of Bielefeld, Germany}

\begin{document}

\section{Introduction}

In heavy ion collisions at ultra-relativistic energies nuclear matter can be studied under extreme conditions. At zero baryo-chemical potential and finite temperature, a cross-over between the hadron gas at low temperatures and the quark gluon plasma at high temperatures has been established by lattice QCD \cite{Borsanyi:2013bia, Bazavov:2014pvz} and also been confirmed by experimental measurements at the Large Hadron Collider (LHC) and the Relativistic Heavy Ion Collider (RHIC) in combination with dynamical models \cite{Pratt:2015zsa}. At finite net baryon densities, current lattice calculations are not applicable and insights about the structure of the QCD phase diagram can only be achieved by experimental exploration in combination with detailed dynamical modeling and phenomenological approaches providing input on the equation of state in the whole temperature ($T$)-baryo-chemical potential ($\mu_B$) plane. 

The previous and currently running programs at CERN-SPS and the recent beam energy scan (BES) program at RHIC provide a comprehensive data set on general bulk observables and selected fluctuation/correlation measurements. In the future, the second generation of experiments with improved capabilities for rare probes will meticulously explore the exciting energy regime accessible at FAIR, NICA, and stage 2 of the BES@RHIC. The second goal, besides obtaining insight on the type of the  phase transition of strongly interacting matter, is to understand the properties of the state of matter that is formed, the QGP. 

One of the most promising observables is the anisotropic flow, that is measured in terms of Fourier coefficients of the azimuthal distribution of produced particles. Directed flow $v_1$ and elliptic flow $v_2$ are analysed extensively to gain insights on the transport properties and the geometry of heavy ion collisions. Within the last 5 years the whole plethora of odd higher coefficients has been studied to investigate the initial state and its fluctuations in more detail. The flow coefficients - as indicated by the name - are a sign of collective behaviour and quantify the response of the system to spatial anisotropies, translated to momentum space by pressure gradients. 

To connect the final state particle distributions to the quantities of interest, e.g. the equation of state  or the viscosity to entropy ratio of QCD matter, detailed dynamical models are necessary. The current state of the art is to use a combination of 3+1 dimensional (viscous) hydrodynamic evolution and hadronic transport for the non-equilibrium evolution in the late stages of the reaction. There are various ways to model/parametrize the initial state for these calculation ranging from Glauber/CGC type models, to classical Yang-Mills evolution, AdS/CFT or transport approaches. At lower beam energies the initial non-equilibrium evolution takes a non-negligible time and one needs to certainly pay attention to the type of initial dynamics that is required. 

In the present study, the calculations are performed within the Ultra-relativistic Quantum Molecular Dynamics (UrQMD) approach including a 3+1 dimensional ideal hydrodynamic evolution where appropriate. The main ingredients of the model are described in Section \ref{hybrid_model}. After that, Section \ref{flow} contains a summary of the expectations for the beam energy dependence of anisotropic flow in heavy ion collisions and the three subsections \ref{elliptic_flow}, \ref{tri_flow} and \ref{directed_flow} are devoted to $v_1$ to $v_3$ in more detail. The last Section \ref{summary} summarizes the main conclusions. 

\section{Hybrid Approach at non zero $\mu_B$}
\label{hybrid_model}
Heavy ion collisions are processes that do not create a system at fixed temperature and density that can be studied over an extended period of time, but rather span a fluctuating finite region in the phase diagram evolving dynamically. Hybrid approaches, based on (viscous) hydrodynamics for the hot and dense stage of the evolution, and non-equilibrium transport for the dilute stages of the reaction are successfully applied to describe bulk observables at high RHIC and LHC energies. These approaches combine the advantages of hydrodynamics and transport and apply both approximations within their respective regions of validity. The equation of state and transport coefficients of hot and dense QCD matter are direct inputs to the hydrodynamic equations, which allows for a controlled modeling of the phase transition. Microscopic transport on the other hand describes all particles and their interactions and provides the full phase-space distribution of produced particles in the final state, which allows for an apples-to-apples comparison to experimental data. 

At higher beam energies the standard picture of the dynamical evolution of a heavy ion reaction consists of non-equilibrium initial state dynamics providing the initial state for the viscous hydrodynamics equations and a hadronic afterburner to account for the separation of chemical and kinetic freeze-out. At lower beam energies, it is not clear how close the produced system really gets to local equilibration and if it thermalizes in the whole phase-space. In general, it is expected that dissipative effects are larger for low beam energies and the hadronic interactions gain importance. In addition, the widely used assumption of boost invariance breaks down and a 3+1 dimensional calculation is inevitable. Also, the equation of state and the transport coefficients are required not only as a function of temperature, but in the whole $T-\mu_B$ - plane. The finite net baryon current needs to be conserved during the evolution. Going beyond the standard hybrid approach by exploring the non-equilibrium dynamics in the vicinity of a critical endpoint or a first order phase transition is the ultimate goal, where exploratory work has been performed so far \cite{Nahrgang:2011mg}. 

The dynamical approach that is used to calculate the results presented below is the Ultra-relativistic Quantum Molecular Dynamics (UrQMD) approach coupled to ideal relativistic fluid dynamics for the hot and dense stage (SHASTA) \cite{Bass:1998ca, Bleicher:1999xi, Petersen:2008dd}. This approach has recently been improved to include the finite shear viscosity during the hydrodynamic evolution as presented in another contribution to this conference and in \cite{Karpenko:2015xea}. The initial state is produced by generating nucleons according to Woods-Saxon distributions and computing dynamically the first binary interactions until the two nuclei have geometrically passed through each other. Event -by-event fluctuations of the positions of the nucleons and of the energy deposition per collision are naturally included. At $t_{\rm start} = \frac{2R}{\gamma v}$ the energy, momentum and net baryon density distributions are calculated by representing the individual particles with a three dimensional Gaussian distribution. Based on these initial conditions including fluctuating velocity profiles the ideal relativistic hydrodynamic equations are solved. There are different options for the equation of state, e.g. a bag model equation of state matched to a hadron gas at low temperatures to explore the sensitivity of observables to a strong first order phase transition (BM) or a more realistic equation of state provided by a chiral model that is fitted to lattice QCD at zero $\mu_B$, reproduces the nuclear ground state properties and incorporates constraints from neutron star properties. At a constant energy density a hypersurface is constructed and particles are sampled according to the Cooper-Frye formula. The hadronic rescattering and decays are treated by the hadronic transport approach. This approach incorporates most of the above mentioned ingredients that are necessary for a realistic dynamical description of heavy ion reactions at lower beam energies. The current strategy can be seen as a first step to take the well-established picture at high energies and explore how well it works at lower beam energies. 

\section{Anisotropic Flow Observables}
\label{flow}

\begin{figure}[hbt]
\centering
\includegraphics[width=7cm]{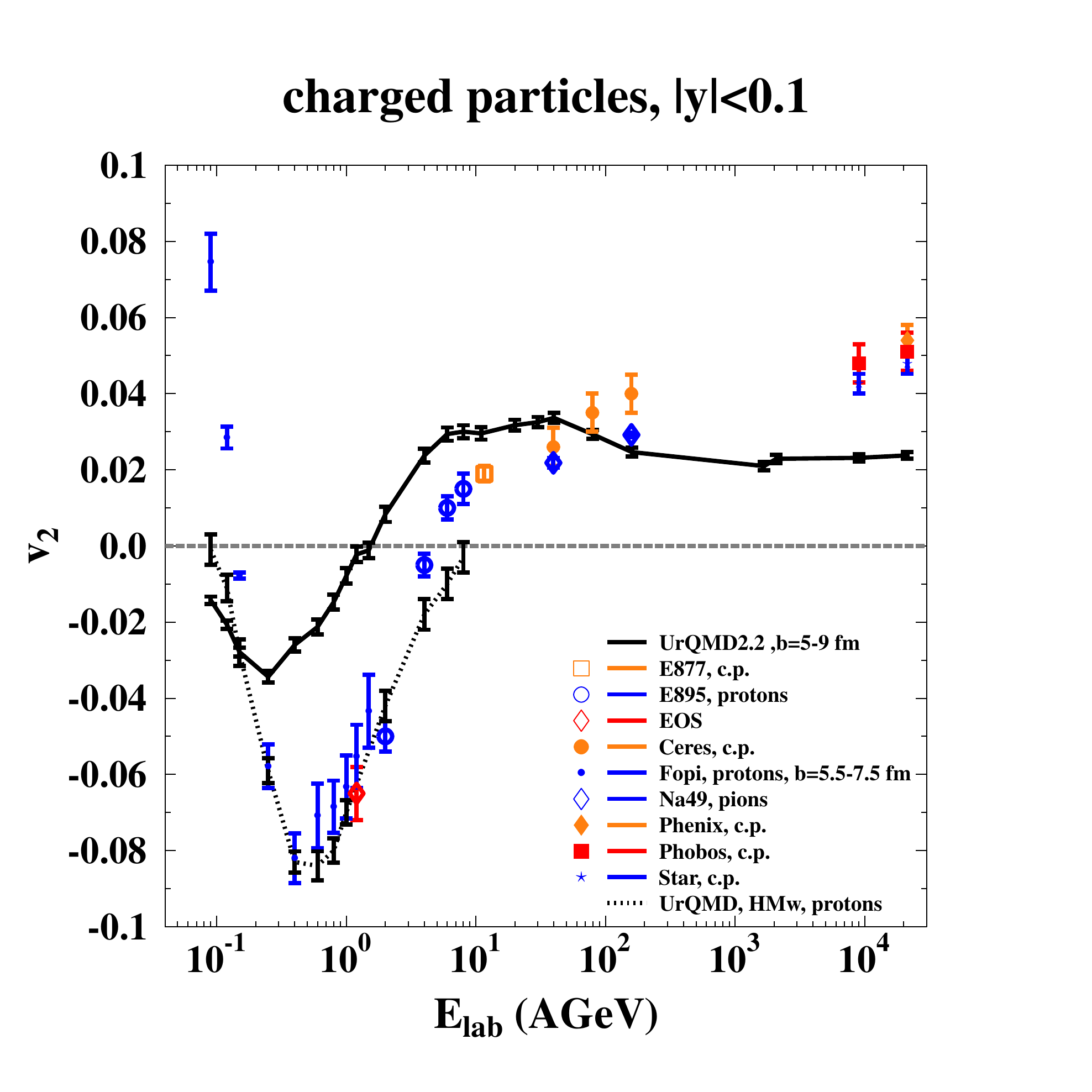}
\includegraphics[width=8cm]{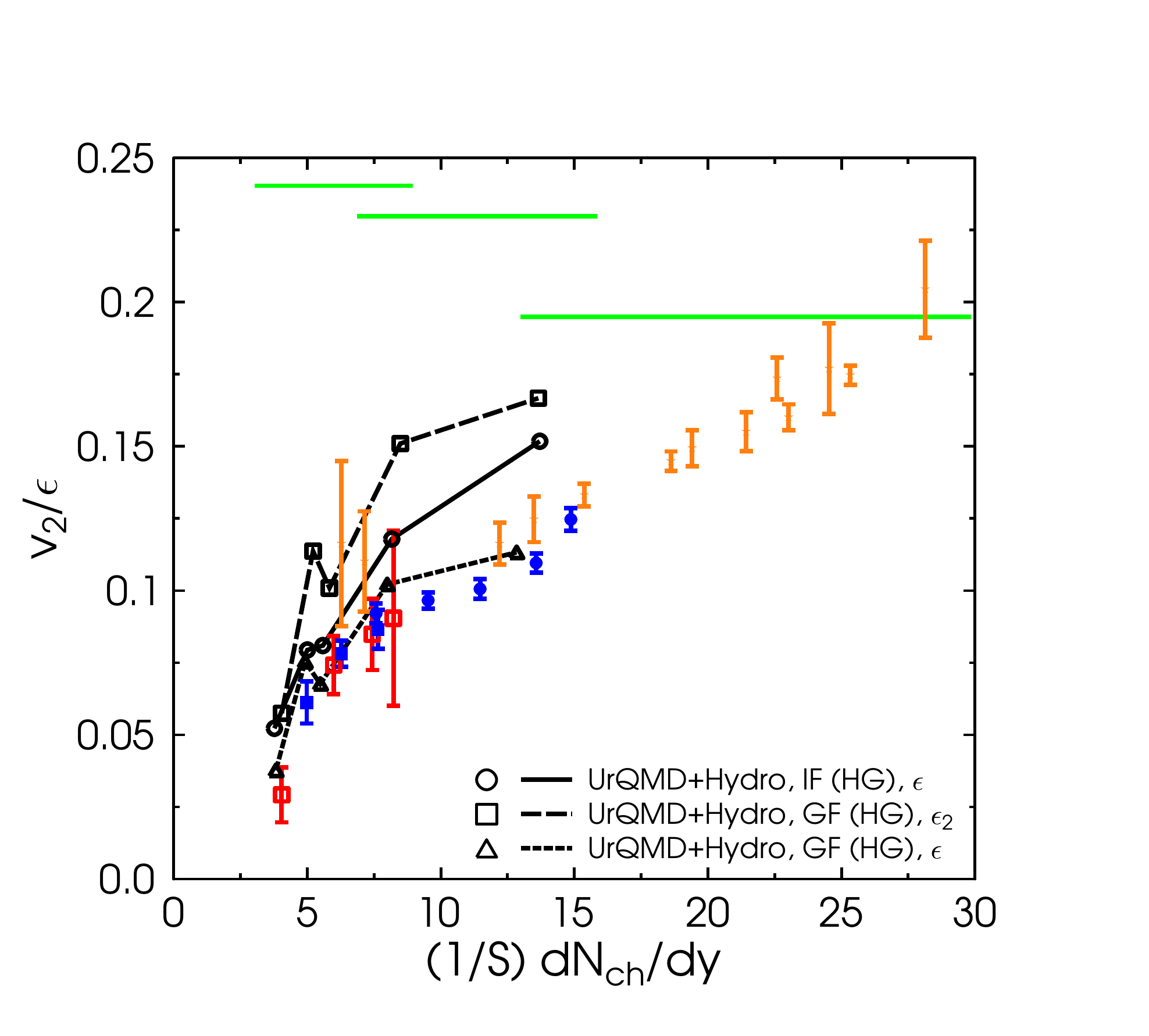}
\caption{(Color online) Left: The calculated beam energy excitation function of elliptic flow of charged particles in Au+Au/Pb+Pb 
collisions in mid-central collisions (b=5-9 fm) with $|y|<0.1$(full line). This curve is compared to data from 
different experiments for mid-central collisions (see \cite{Petersen:2009vx} for refs). The dotted line 
in the low energy regime depicts UrQMD calculations with the mean field \cite{Li:2006ez}. Fig. from \cite{Petersen:2006vm}. Right: $v_2/\epsilon$ as a function of $(1/S)dN_{\rm ch}/dy$ for different energies and centralities for Pb+Pb/Au+Au collisions compared to data \cite{Alt:2003ab}. The results from mid-central collisions (b=5-9 fm) calculated within the hybrid model with different freeze-out transitions and different definitions of the eccentricity are shown by black lines. The green full lines correspond to the previously calculated hydrodynamic limits \cite{Kolb:2000sd}. Fig. taken from \cite{Petersen:2009vx}.}
\label{figv2exc}
\end{figure}

Anisotropic flow is the collective response to geometrical structures in the initial state distribution of heavy ion collisions. If there are strong enough interactions spatial anisotropies are getting translated to momentum space throughout the evolution. Therefore, anisotropic flow is supposed to be very sensitive to the equation of state and to the transport properties of the produced matter. It is quantified by the Fourier coefficients of the azimuthal distribution of the produced particles 
\begin{equation}
v_n = \cos (n\phi -\Psi_n)
\end{equation} 
where $\Psi_n$ is the event plane and $n$ specifies the coefficient of interest. There are many different two-particle and many-particle methods to measure the anisotropic flow coefficients with different sensitivities to non-flow and fluctuations. In this article, either the event plane method (some $v_2$ results and $v_3$) or the theoretical definition with respect to the known reaction plane ($v_1 = \langle \frac{p_x}{p_x^2+p_y^2}\rangle$ and $v_2 = \langle \frac{p_x^2-p_y^2}{p_x^2+p_y^2}\rangle$) have been applied to calculate the flow observables.

Elliptic flow is one of the main observables to support the claim that the quark gluon plasma behaves like a nearly perfect liquid. Therefore, the beam energy dependence of elliptic flow is supposed to be sensitive to the phase transition and the changing transport properties at lower collision energies. Triangular flow is the first odd flow component that is mainly sensitive to fluctuations and would be zero, if the initial state is taken as average over many events. Therefore, the hope is that measuring the beam energy dependence of triangular flow helps to disentangle the 'trivial' initial state fluctuations from the interesting fluctuations that arise due to the critical dynamics. Last but not least directed flow is first of all a measure of the initial angular momentum in the heavy ion collision. The beam energy dependence was predicted to show a dip structure when a first order phase transition occurs during the evolution. In the following Sections, the beam energy dependence of each of these observables will be discussed in more detail. 

\begin{figure}
\centering
\includegraphics[width=6.5cm]{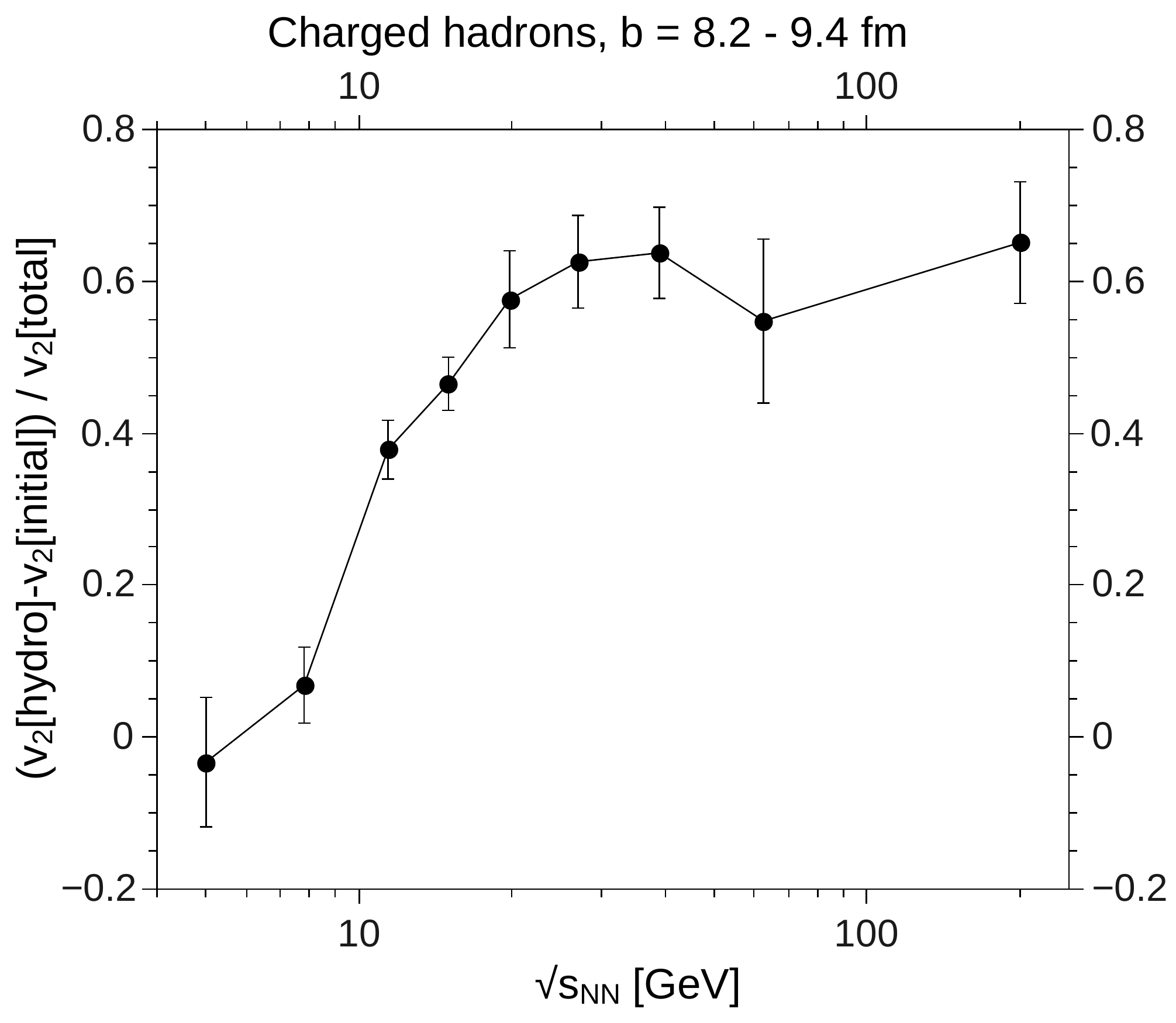}
\includegraphics[width=6.5cm]{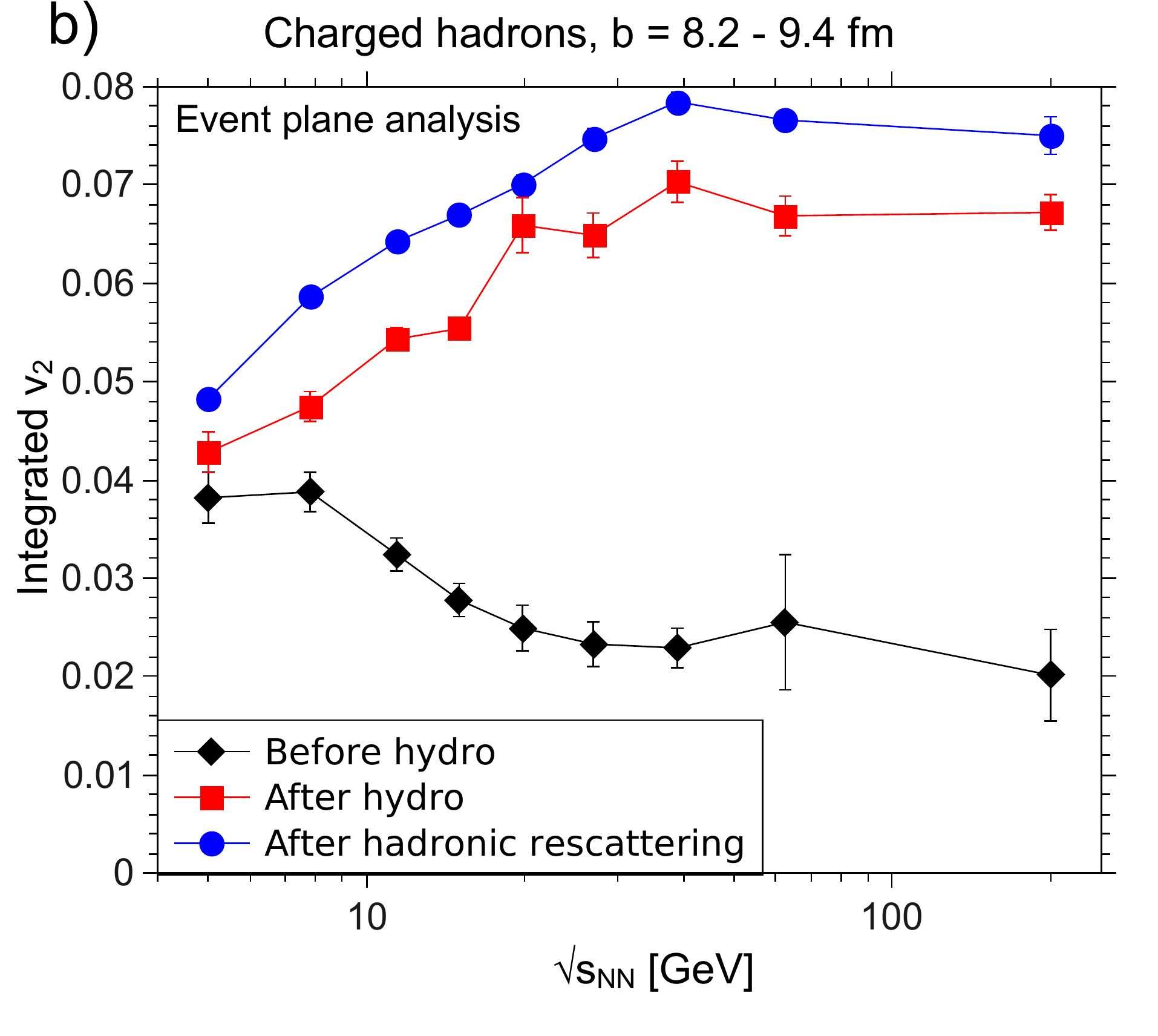}
\caption{(Color online) Left: Fraction of elliptic flow generated during hydrodynamic evolution as a function of beam energy. Right: Magnitude of $v_2\{\textrm{EP}\}$ at the beginning of 
hydrodynamical evolution (squares), immediately after particlization (diamonds) and after 
the full simulation (circles) in midcentral collisions. Fig. from \cite{Auvinen:2013sba}.}
\label{figv2phases}
\end{figure}

\subsection{Elliptic flow - a measure of the perfect fluid}
\label{elliptic_flow}

\begin{figure}[hb]
\centering
\includegraphics[width=7cm]{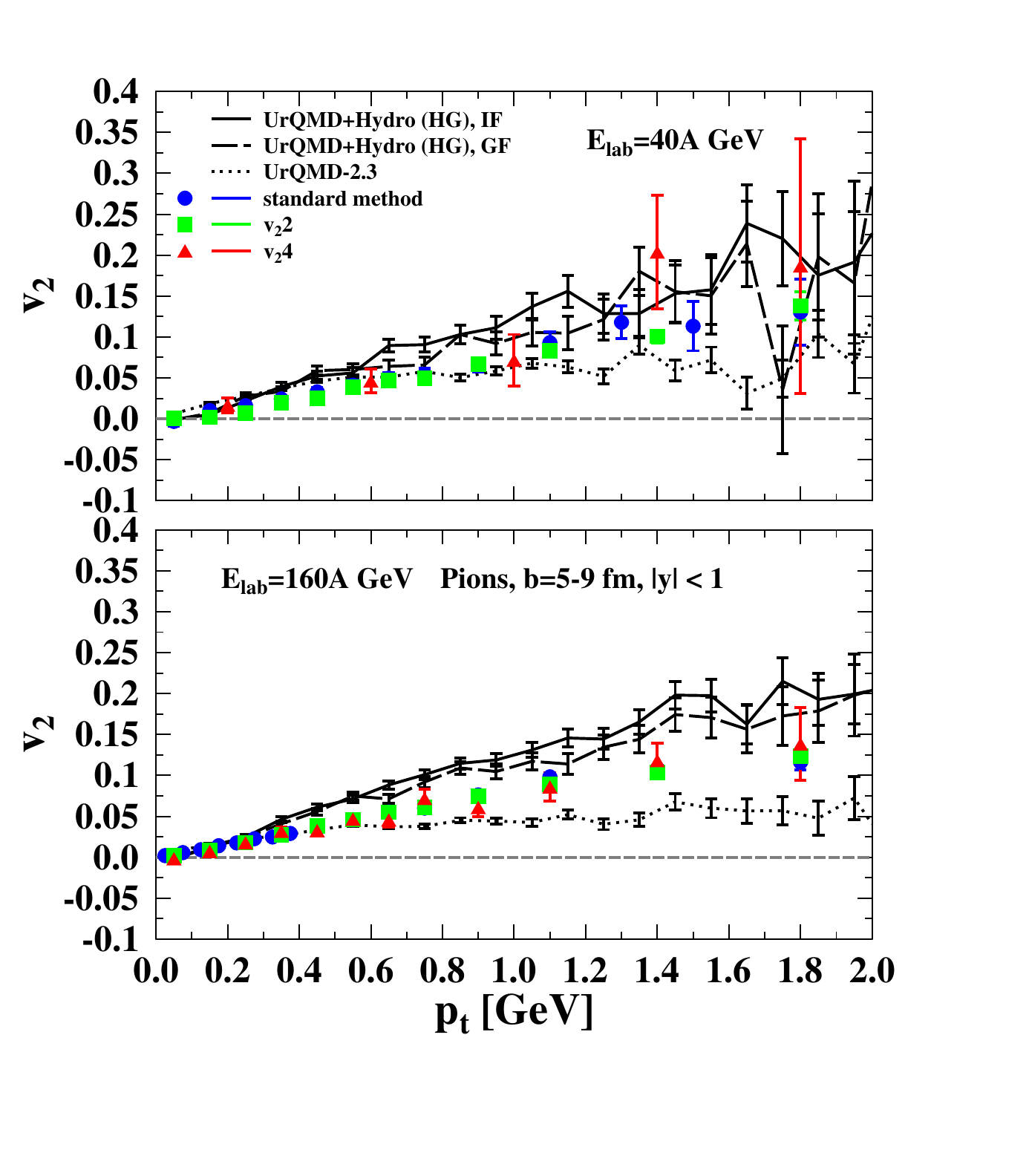}
\includegraphics[width= 7cm]{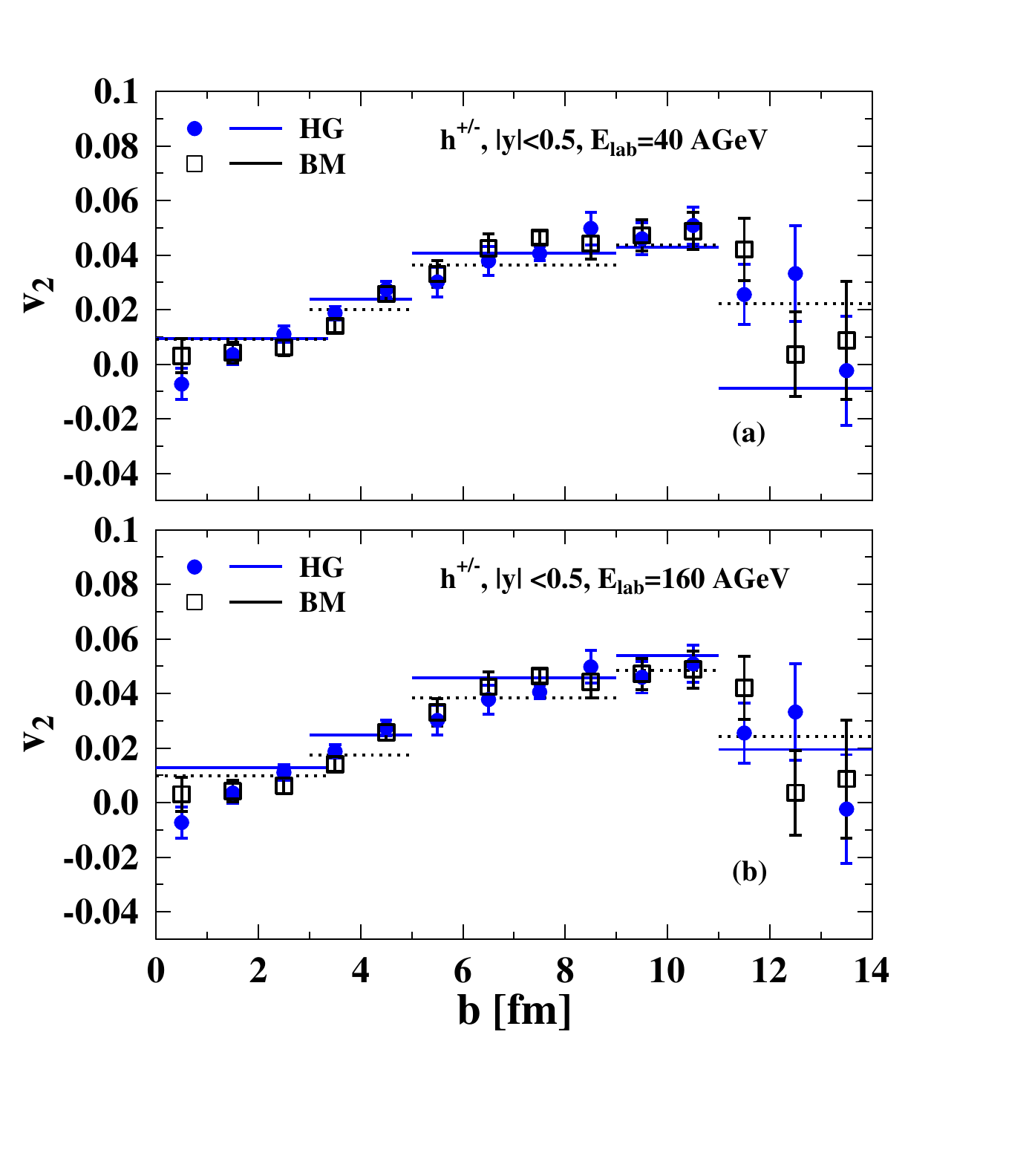}
\caption{(Color online) Left: Elliptic flow of pions in mid-central (b=5-9 fm) Pb+Pb collisions at $E_{\rm lab}=40A~$GeV and  $E_{\rm lab}=160A~$GeV. The full and dashed black lines depict the hybrid model calculation, while the pure transport calculation is shown as the black dotted line. The colored symbols display experimental data obtained with different measurement methods by NA49 \cite{Alt:2003ab}. Fig. taken from \cite{Petersen:2009vx}.
Right: Centrality dependence of elliptic flow of charged
particles at midrapidity ($|y|<0.5$) for Pb+Pb collisions at $E_{\rm lab}=40A$
GeV (a) and $E_{\rm lab}=160A$ GeV (b). The horizontal lines indicate the
results for averaged initial conditions using the hadron gas EoS (blue full
line) and the bag model EoS (black dotted line) while the symbols (full circles for
HG-EoS and open squares for BM-EoS) depict the results for the event-by-event
calculation. Fig. taken from \cite{Petersen:2010md}.}
\label{figv2_ptpi}      
\end{figure}
The second Fourier coefficient, the so called elliptic flow, is the response of the system to the initial almond shaped overlap region in non-central collisions. Due to fluctuations $v_2$ is also non-zero in central collisions, but it is mainly generated due to the overall shape deformation of the fireball. Only if the mean free path is small enough the initial spatial anisotropy is converted to the corresponding momentum space anisotropy. The beam energy dependence of elliptic flow is quite interesting. At very low energies, the spectator nuclei are blocking the interaction region and the so called 'squeeze-out' leads to negative elliptic flow values with respect to the reaction plane. In this region the nuclear interactions are important and a hadron transport approach including mean fields can describe the elliptic flow rather well (see Fig. \ref{figv2exc}, left). At high beam energies the elliptic flow turns positive and grows as a function of beam energy. In the intermediate region around 10-40A GeV the hadron transport approach reaches the right "ball park" values while the underestimation at high beam energies is a sign for the importance of partonic interactions starting at around 160A GeV. 

Fig. \ref{figv2exc} (right) shows the response function $v_2/\epsilon_2$ as a function of the charged particle density. The green lines indicate the expectations from ideal hydrodynamic calculations and that this line meets the experimental data at $\sqrt{s_{\rm NN}} =  200$ GeV was the basis for the claim that a perfect fluid has been created. The black lines show various UrQMD hybrid calculations to demonstrate that the generic behaviour of this curve can be reproduced once the non-equilibrium stages of the reaction are treated properly.

To investigate how this energy dependence of elliptic flow is generated in the hybrid approach the contribution to the final integrated elliptic flow of charged particles has been calculated and the percentage is shown in Fig. \ref{figv2phases} (left). At lower energies, the hydrodynamic evolution lasts only a few fm/c, therefore almost all the elliptic flow is generated by the hadronic transport approach. At higher energies the hydrodynamic evolution is necessary to reach the high elliptic flow values as discussed above. Fig. \ref{figv2phases} (right) shows that the contribution from the late hadronic rescattering is constant at about 10 \% throughout the whole energy regime investigated. 

Fig. \ref{figv2_ptpi} (left) shows the transverse momentum dependent elliptic flow of pions. NA49 measurements are compared to hybrid and pure hadronic transport calculations and again it is clearly visible that at 40 AGeV the hydrodynamic evolution does not play a role while at 160 AGeV the hydrodynamic evolution is crucial to reproduce the elliptic flow. The dependence on initial state fluctuations and the equation of state has been studied in Fig. \ref{figv2_ptpi} (right), where the centrality dependence of elliptic flow of charged particles is shown. There is no difference visible between the fluctuating events (symbols) and the averaged smooth initial conditions (horizontal lines) as expected for elliptic flow. The difference between the bag model equation of state with a strong first order phase transition and the hadron gas equation of state is not observable either.

\subsection{Triangular flow - a measure of fluctuations}
\label{tri_flow}
\begin{figure}
\centering
\includegraphics[width=6.5cm]{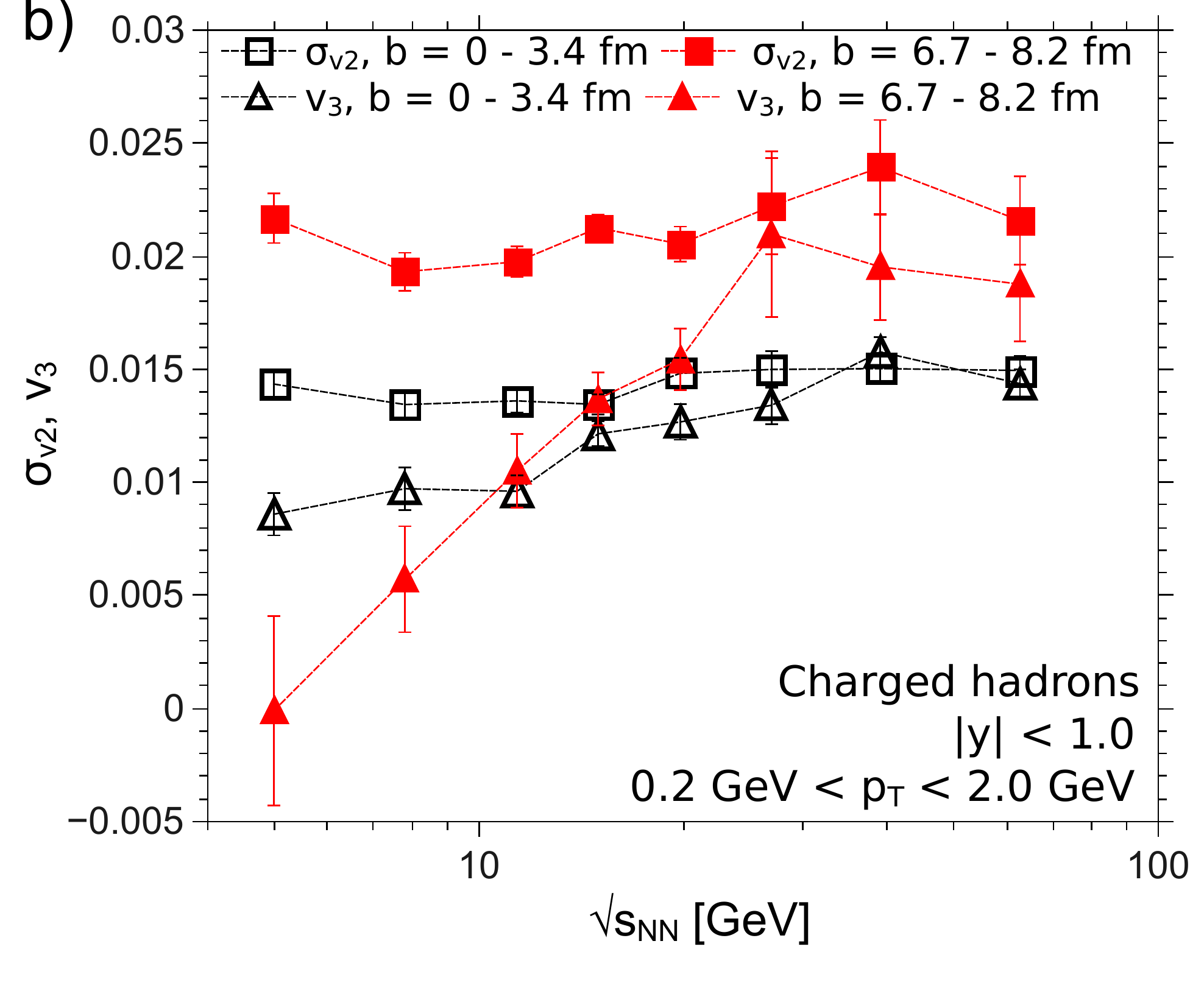}
\includegraphics[width=6.5cm]{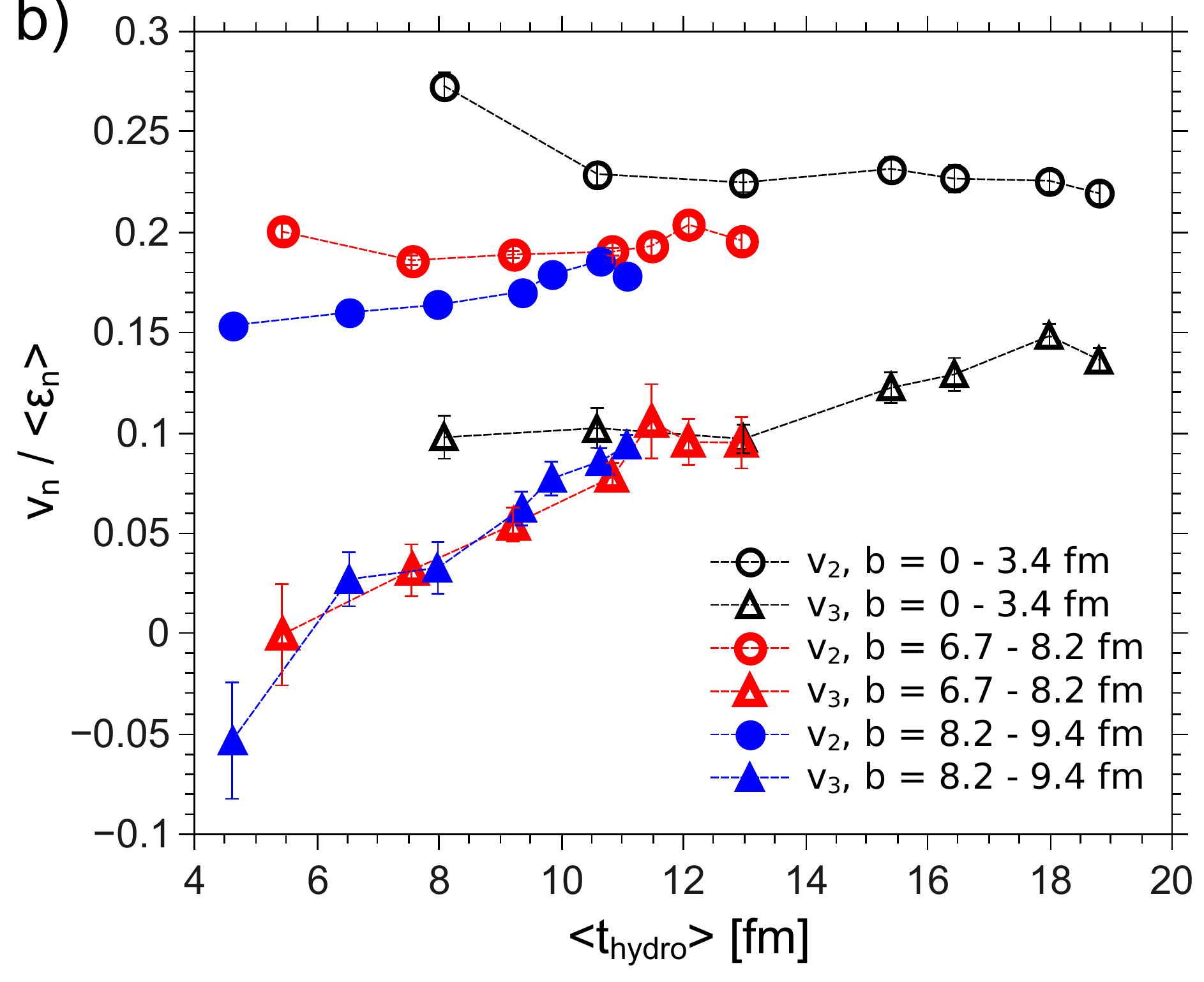}	
\caption{(Color online) Left: $v_3\{\textrm{EP}\}$ compared with 
initial state fluctuations' contribution to $v_2$,
$\sigma_{v2}=\sqrt{\frac{1}{2}(v_2\{\textrm{EP}\}^2-v_2\{\textrm{RP}\}^2)}$ 
(squares). Right: Scaled flow coefficients $v_2/ \langle \epsilon_2 \rangle$ and 
$v_3/ \langle \epsilon_3 \rangle$ with respect to the 
average total hydro duration for impact parameter ranges $b = 0-3.4$ fm, $6.7-8.2$ fm 
and $8.2-9.4$ fm. Figs taken from \cite{Auvinen:2013sba}.}
\label{fig_v3}
\end{figure}

Since 2010 event by event fluctuations in the initial state have been studied as the source of higher odd anisotropic flow coefficients \cite{Adare:2012kf, Luzum:2013yya}. These odd coefficients average to zero without event by event fluctuations. To demonstrate that $v_3$ is really sensitive to fluctuations the beam energy dependence of triangular flow of charged particles is compared to the dynamical fluctuations of elliptic flow $\sigma_{v_2}$ in Fig. \ref{fig_v3} (left). In central collisions both have roughly the same value whereas in non-central collisions they have a similar magnitude only at high energies. At lower beam energies triangular flow decreases. 

If non-zero triangular flow is measured at lower beam energies and it is generated by initial spatial fluctuations it serves as a nice probe to disentangle these 'trivial' fluctuations from more interesting fluctuations due to the phase transition or critical endpoint. This option relies of course on the fact that a finite triangular flow needs to be present. As Fig. \ref{fig_v3} (right) shows, $v_3$ scales for different energies and centralities with the duration of the hydrodynamic evolution in the hybrid approach. The circles show that for elliptic flow the influence of the hadronic non-equilibrium evolution is much larger. Therefore, triangular flow is much more sensitive to a finite viscosity and might even vanish at lower beam energies. On the positive side, this disappearance of triangular flow can be seen as a more direct evidence that the perfect fluid QGP does not persist for a long enough time. 

\subsection{Directed flow - a measure of the phase transition}
\label{directed_flow}

Last but not least, let us come to one more anisotropic flow observable. The directed flow measures the stream of particles within the reaction plane and has opposite sign in the forward and backward hemisphere. Nowadays, a rapidity-even $v_1$ dipole component has also been observed, but here we concentrate on the traditional rapidity odd measure. To plot the beam energy dependence one extracts the slope of directed flow at midrapidity.

\begin{figure}[t]	
\includegraphics[width=6.5cm]{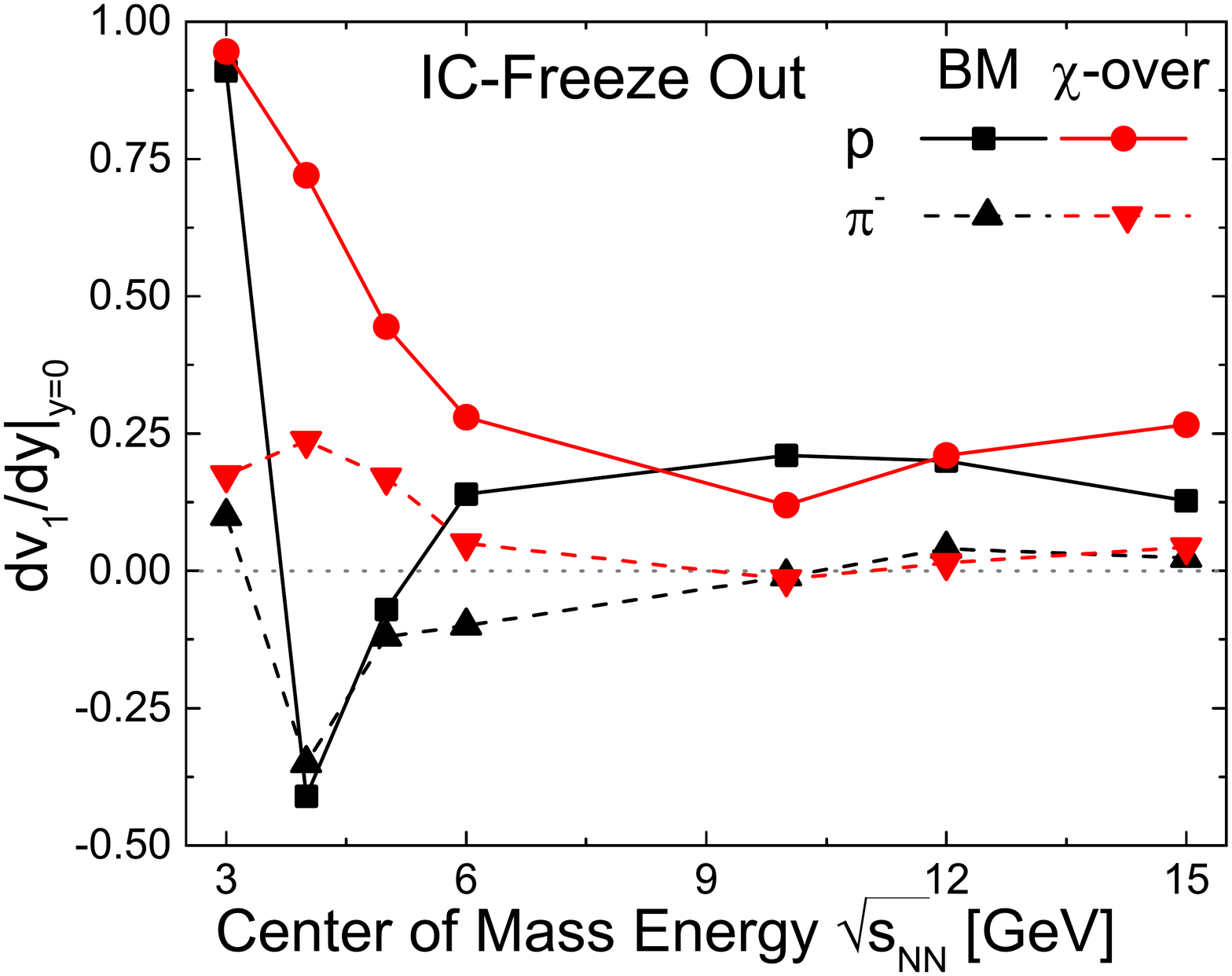}	
\includegraphics[width=6.5cm]{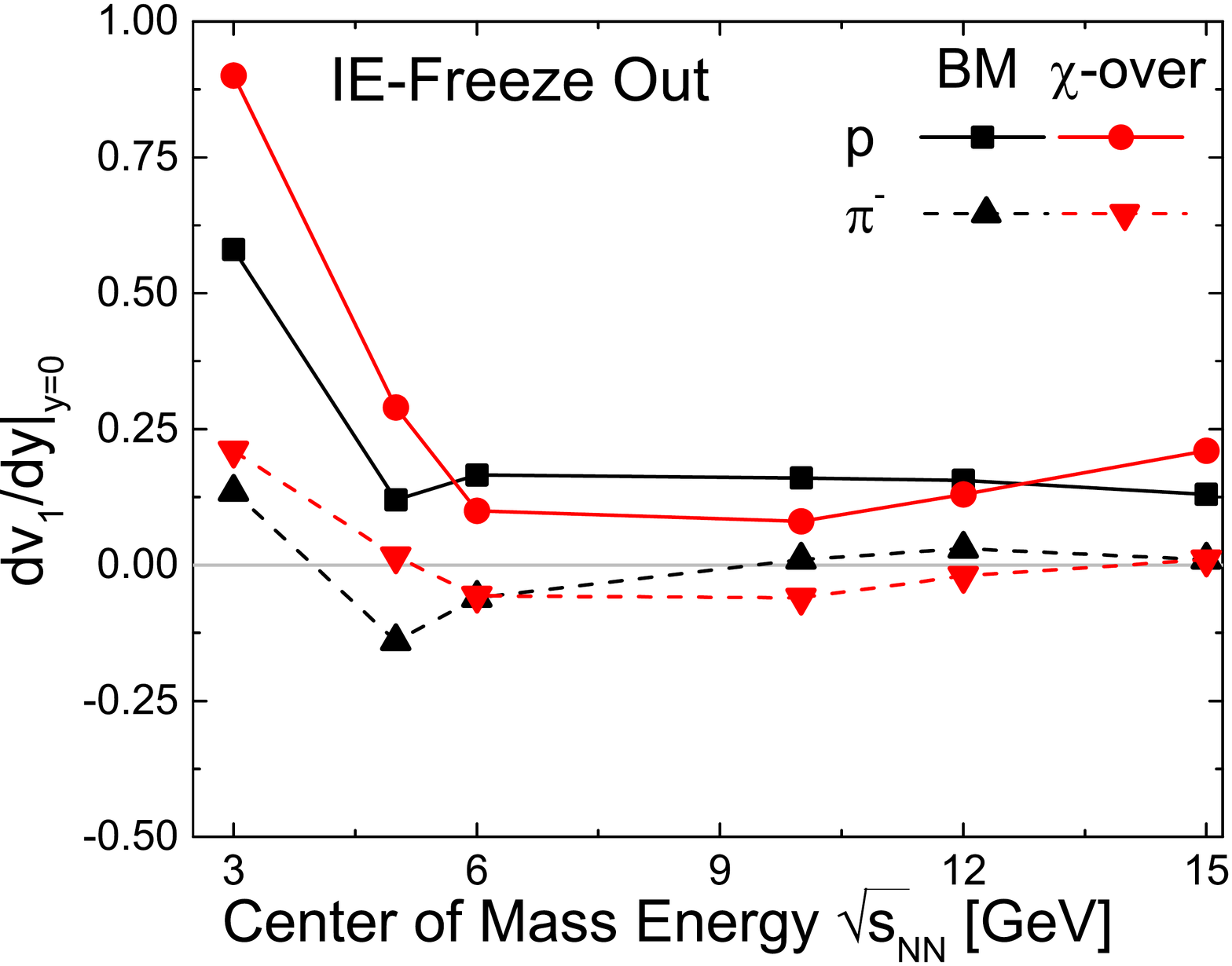}	
\caption{[Color online] Left: Slope of $v_1$ of protons and pions around mid rapidity extracted from the ideal 1-fluid calculations with a bag model and crossover EoS. For particle production we applied a Cooper-Frye prescription on a iso-chronous hypersurface. Right: Slope of $v_1$ of protons and pions around mid rapidity extracted from the ideal 1-fluid calculations with a bag model and crossover EoS. For particle production we applied a Cooper-Frye prescription on a iso-energy density hypersurface. Figs. taken from \cite{Steinheimer:2014pfa}.
}\label{fig_fluid}
\end{figure}	

Based on the AGS and NA49 measurements there have been predicitions within hydrodynamic calculations that the slope of the directed flow of protons has a dip as a function of beam energy, signaling the phase transition to the quark gluon plasma \cite{Brachmann:1999xt, Csernai:1999nf}. In light of the new STAR measurements from the RHIC beam energy scan program \cite{Adamczyk:2014ipa}, this old observable has been recalculated with modern theoretical techniques. In Fig. \ref{fig_fluid} a one fluid calculation has been performed. This resembles the previous predictions enhanced by actual particle sampling and a $v_1$ calculation that can be compared to experimental measurements. On the left hand side, the rather unphysical scenario of a constant time freeze-out is shown, whereas the right hand side corresponds to a more realistic iso-energy density transition scenario. The two figures demonstrate, that the freeze-out dynamics has a significant effect on the structures in the excitation function of the slope of the directed flow of protons and pions. The large difference between the two different scenarios  - with and without first order phase transition - disappears. 

\begin{figure}[t]	
\includegraphics[width=6.5cm]{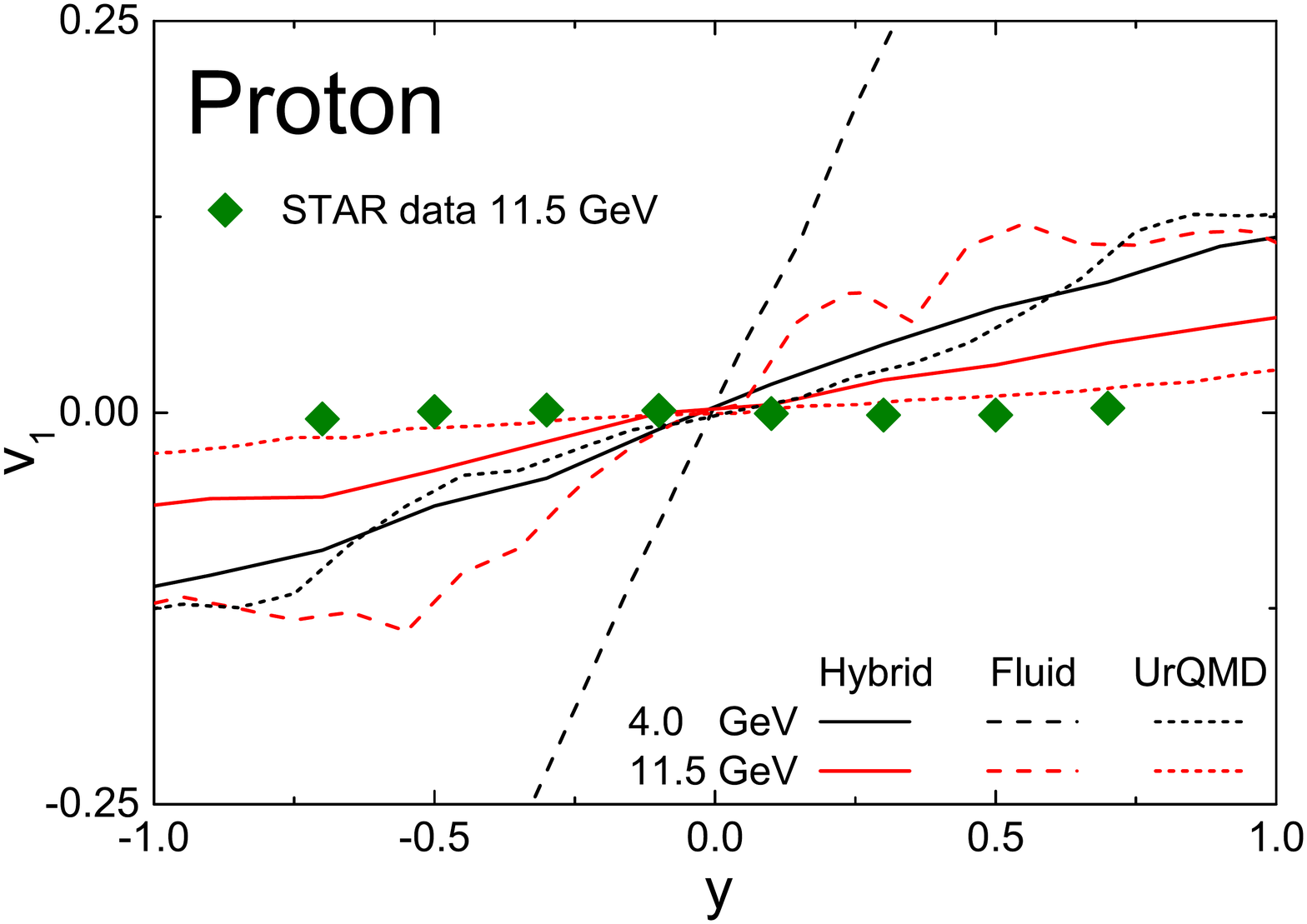}
\includegraphics[width=6.5cm]{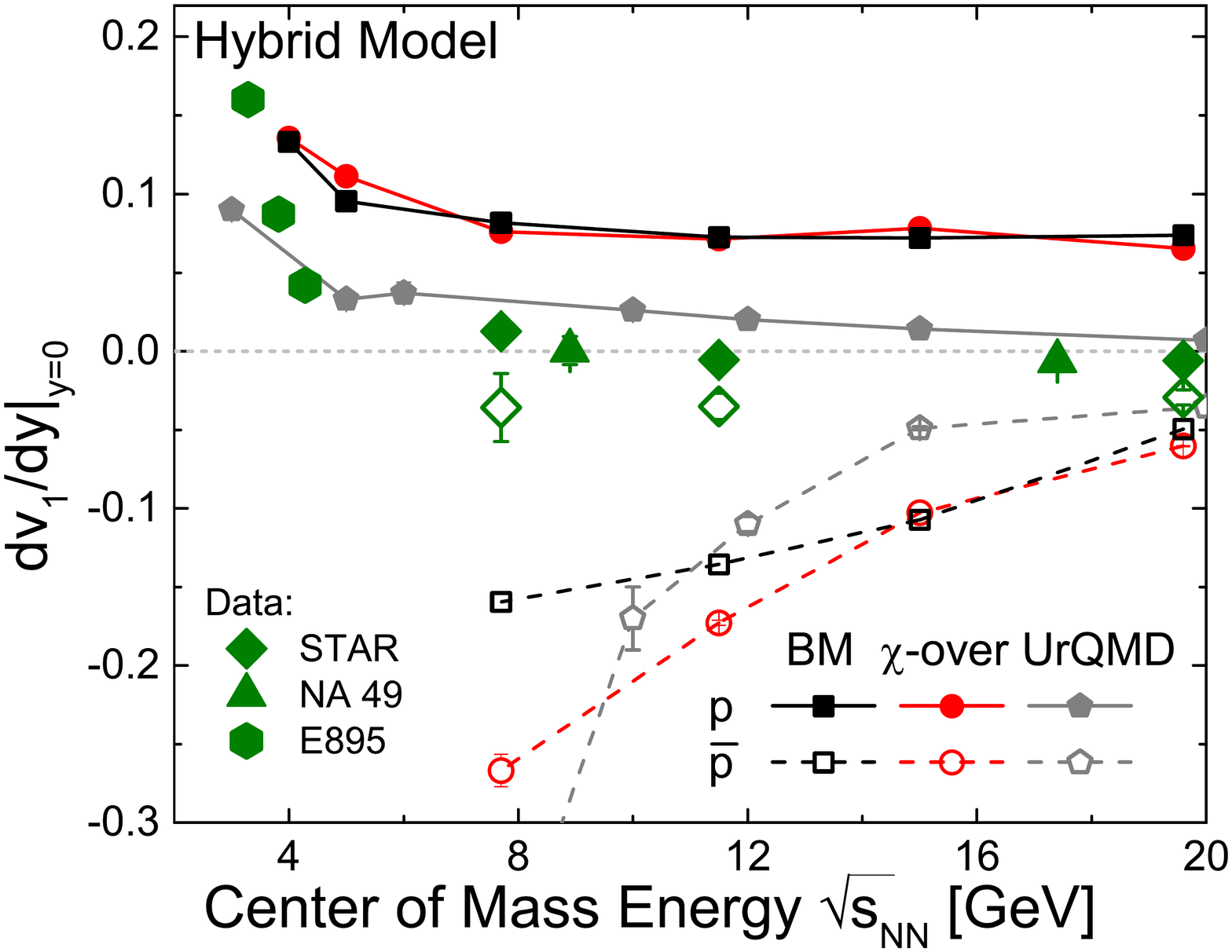}	
\caption{[Color online] Left: Comparison of proton $v_1(y)$ for the different model applied. Right: Slope of $v_1$ of protons and anti-protons around mid rapidity extracted from the hybrid model calculations with a bag model and crossover EoS. We compare with standard UrQMD and experimental data.Figs. taken from \cite{Steinheimer:2014pfa}.
}\label{fig_v1proton}
\end{figure}		

\begin{figure}[t]	
\includegraphics[width=6.5cm]{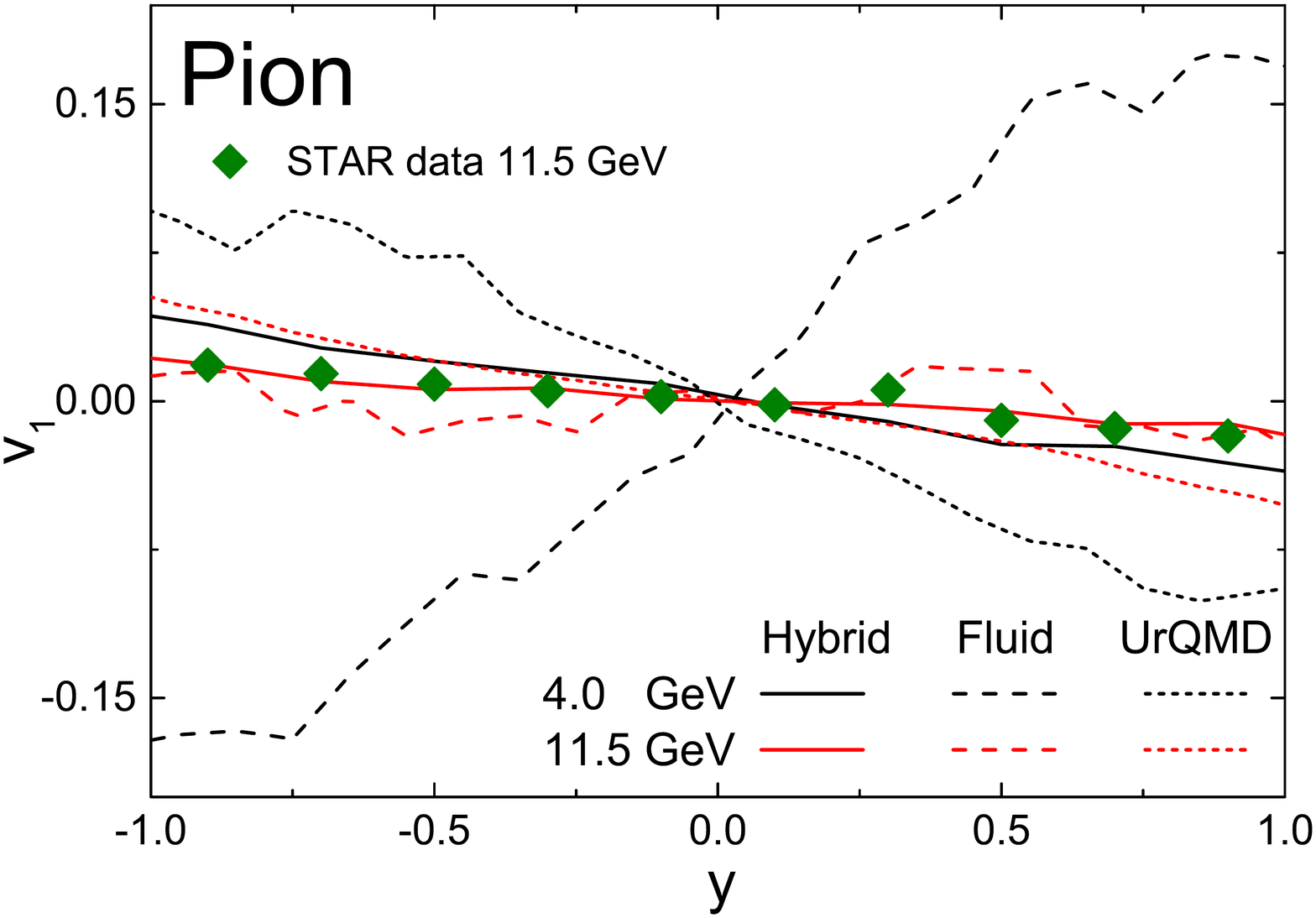}	
\includegraphics[width=6.5cm]{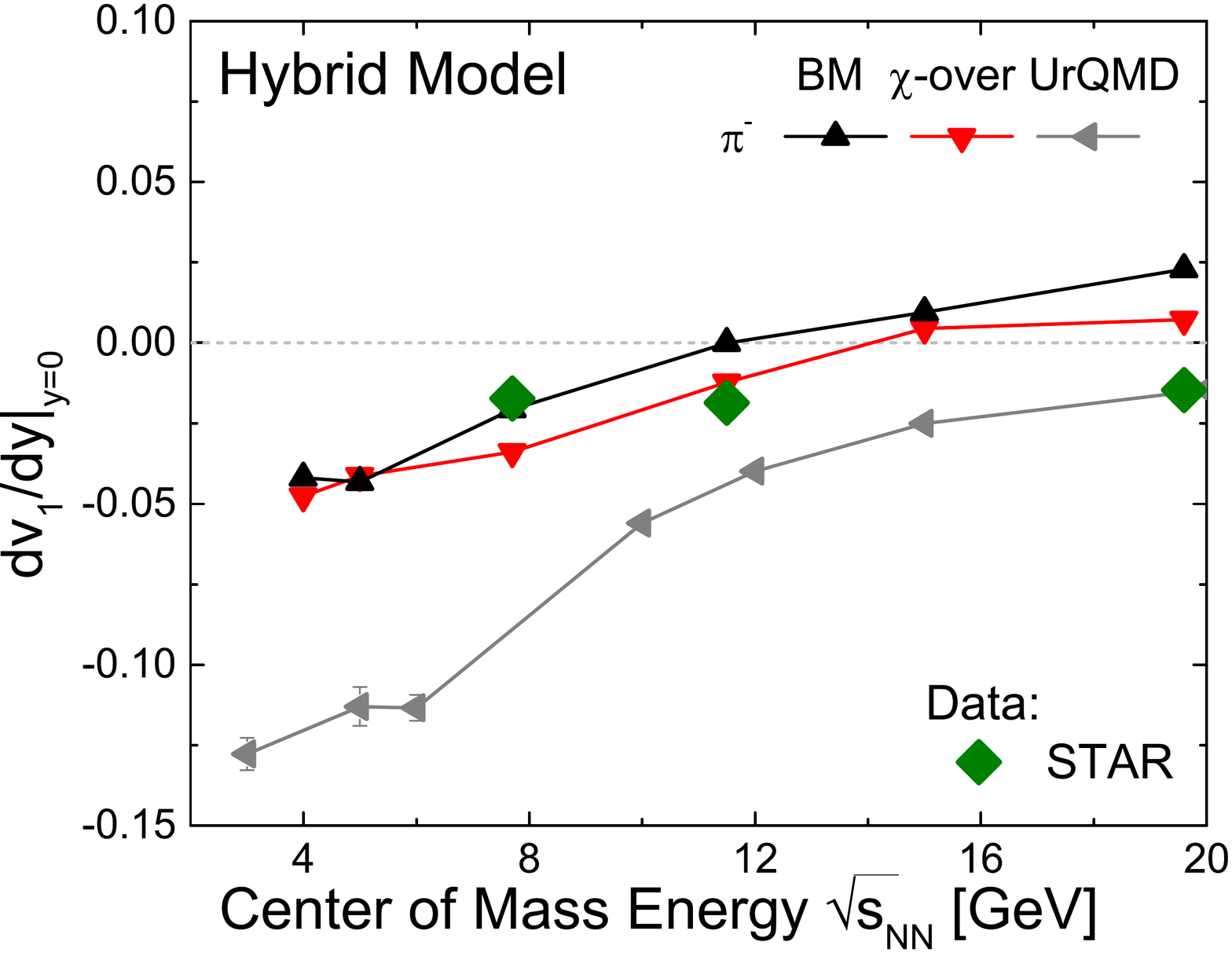}	
\caption{[Color online] Left: Comparison of pion $v_1(y)$ for the different model applied. Right: Slope of $v_1$ of negatively charged pions around mid rapidity extracted from the hybrid model calculations with a bag model and crossover EoS. We compare with standard UrQMD and experimental data.Figs. taken from \cite{Steinheimer:2014pfa}.
}\label{fig_v1pion}
\end{figure}

Figs. \ref{fig_v1proton} and \ref{fig_v1pion} both show on the left hand side the directed flow as a function of rapidity at two different beam energies and on the right hand side the slope around mid rapidity that has been extracted to display the full beam energy dependence. Especially for the proton directed flow it is clear that the signal is very small around zero. Even though all the different calculations capture the positive slope it is obvious that pure one fluid calculations lead to too much directed flow, while surprisingly the pure hadron transport shows the best agreement with experimental data at this point. 
  
For pions the negative slope arises because of a shadowing effect. The newly produced particles are blocked by the spectators and therefore the slope of directed flow is negative at low beam energies. Again the fluid calculation overestimates the flow at low beam energies and for pions the hybrid UrQMD approach seems to describe the measurements rather well. More important than a quantitative comparison to experimental data is the conclusion that the hybrid calculations with different assumptions about the order of the phase transition do not show any sensitivity on these observables. Detailed measurements of the centrality dependence of directed flow are going to allow for better conclusions on the stopping mechanism, the nuclear mean fields and the equation of state that are all crucial to understand the dynamics of heavy ion collisions at low beam energies.  

\section{Summary and Conclusions}
\label{summary}
Collective flow is one of the main observables to understand the properties of the quark gluon plasma and the dynamics of heavy ion reactions. Especially the beam energy dependence of anisotropic flow provides high potential for insights on the equation of state and the dependence of transport coefficients on the baryo-chemical potential. To connect the final state observables to the quantities of interest, hybrid approaches based on relativistic hydrodynamics and non-equilibrium transport provide a realistic description of the dynamics. The main conclusions are that the small change of elliptic flow as a function of the beam energy can be explained by the transport dynamics that gains importance at lower beam energies and replaces the diminished hydrodynamic evolution. Triangular flow is much more sensitive to the viscosity and cannot be generated by hadronic transport alone. The beam energy dependence of the slope of the directed flow around midrapidity has been studied and even though the old predictions are reproduced, there is no difference between a modern hybrid calculation with and without a strong first order phase transition. To reach final conclusions on this subject, more detailed studies including nuclear potentials and an improved treatment of the baryon stopping are necessary. 

\section{Acknowledgements}
H. P. acknowledges funding of a Helmholtz Young Investigator Group VH-NG-822 from
the Helmholtz Association and GSI. This work was supported by the Helmholtz International
Center for the Facility for Antiproton and Ion Research (HIC for FAIR) within the
framework of the Landes-Offensive zur Entwicklung Wissenschaftlich-Oekonomischer Exzellenz (LOEWE) program launched by the State of Hesse. Computational resources have been provided by the Center
for Scientific Computing (CSC) at the Goethe-University of Frankfurt.


\begin{thebibliography}{99}
\bibitem{Borsanyi:2013bia}
  S.~Borsanyi, Z.~Fodor, C.~Hoelbling, S.~D.~Katz, S.~Krieg and K.~K.~Szabo,
  Phys.\ Lett.\ B {\bf 730} (2014) 99.



\bibitem{Bazavov:2014pvz}
  A.~Bazavov {\it et al.}  [HotQCD Collaboration],
  Phys.\ Rev.\ D {\bf 90} (2014) 9,  094503.



\bibitem{Pratt:2015zsa}
  S.~Pratt, E.~Sangaline, P.~Sorensen and H.~Wang,
  arXiv:1501.04042 [nucl-th].


\bibitem{Nahrgang:2011mg}
  M.~Nahrgang, S.~Leupold, C.~Herold and M.~Bleicher,
  Phys.\ Rev.\ C {\bf 84} (2011) 024912.



\bibitem{Bass:1998ca}
  S.~A.~Bass, M.~Belkacem, M.~Bleicher, M.~Brandstetter, L.~Bravina, C.~Ernst, L.~Gerland and M.~Hofmann {\it et al.},
  Prog.\ Part.\ Nucl.\ Phys.\  {\bf 41} (1998) 255
   [Prog.\ Part.\ Nucl.\ Phys.\  {\bf 41} (1998) 225].



\bibitem{Bleicher:1999xi}
  M.~Bleicher, E.~Zabrodin, C.~Spieles, S.~A.~Bass, C.~Ernst, S.~Soff, L.~Bravina and M.~Belkacem {\it et al.},
  J.\ Phys.\ G {\bf 25} (1999) 1859.



\bibitem{Petersen:2008dd}
  H.~Petersen, J.~Steinheimer, G.~Burau, M.~Bleicher and H.~Stocker,
  Phys.\ Rev.\ C {\bf 78} (2008) 044901.



\bibitem{Karpenko:2015xea}
  I.~A.~Karpenko, P.~Huovinen, H.~Petersen and M.~Bleicher,
  arXiv:1502.01978 [nucl-th].


\bibitem{Li:2006ez}
  Q.~f.~Li, Z.~x.~Li, S.~Soff, M.~Bleicher and H.~Stoecker,
  J.\ Phys.\ G {\bf 32} (2006) 407.



\bibitem{Petersen:2006vm}
  H.~Petersen, Q.~Li, X.~Zhu and M.~Bleicher,
  Phys.\ Rev.\ C {\bf 74} (2006) 064908.



\bibitem{Kolb:2000sd}
  P.~F.~Kolb, J.~Sollfrank and U.~W.~Heinz,
  Phys.\ Rev.\ C {\bf 62} (2000) 054909.



\bibitem{Petersen:2009vx}
  H.~Petersen and M.~Bleicher,
  Phys.\ Rev.\ C {\bf 79} (2009) 054904.



\bibitem{Auvinen:2013sba}
  J.~Auvinen and H.~Petersen,
  Phys.\ Rev.\ C {\bf 88} (2013) 6,  064908.

\bibitem{Alt:2003ab}
  C.~Alt {\it et al.}  [NA49 Collaboration],
  Phys.\ Rev.\ C {\bf 68} (2003) 034903.


\bibitem{Petersen:2010md}
  H.~Petersen and M.~Bleicher,
  Phys.\ Rev.\ C {\bf 81} (2010) 044906.



\bibitem{Adare:2012kf}
  A.~Adare, M.~Luzum and H.~Petersen,
  Phys.\ Scripta {\bf 87} (2013) 048001
   [Phys.\ Scripta {\bf 04} (2013) 048001].



\bibitem{Luzum:2013yya}
  M.~Luzum and H.~Petersen,
  J.\ Phys.\ G {\bf 41} (2014) 063102.



\bibitem{Steinheimer:2014pfa}
  J.~Steinheimer, J.~Auvinen, H.~Petersen, M.~Bleicher and H.~St\"ocker,
  Phys.\ Rev.\ C {\bf 89} (2014) 5,  054913.



\bibitem{Brachmann:1999xt}
  J.~Brachmann, S.~Soff, A.~Dumitru, H.~Stoecker, J.~A.~Maruhn, W.~Greiner, L.~V.~Bravina and D.~H.~Rischke,
  Phys.\ Rev.\ C {\bf 61} (2000) 024909.



\bibitem{Csernai:1999nf}
  L.~P.~Csernai and D.~Rohrich,
  Phys.\ Lett.\ B {\bf 458} (1999) 454.

\bibitem{Adamczyk:2014ipa}
  L.~Adamczyk {\it et al.}  [STAR Collaboration],
  Phys.\ Rev.\ Lett.\  {\bf 112} (2014) 16,  162301.


\end{thebibliography}
\end{document}